\documentclass[twocolumn,prd,floatfix,superscriptaddress]{revtex4-2}

\usepackage{physics}
\usepackage{amsmath}
\usepackage{amssymb}
\usepackage{graphicx}

\begin{document}
	
	\title{Optimal Interferometer Geometry for Gravitationally Induced Quantum Entanglement}
	
	\author{Alireza Maleki}
	\affiliation{School of Physics, Institute for Research in Fundamental Sciences (IPM), Tehran, Iran}
	
	\date{\today}
	
	\begin{abstract}
		The quest for a quantum theory of gravity remains a central challenge in modern physics, driven by the goal of unifying general relativity and quantum mechanics. Direct experimental access to quantum gravitational effects is notoriously difficult due to the inaccessibility of the Planck scale. Recent proposals, however, suggest that gravitationally induced quantum entanglement between two massive particles, realized through advanced quantum interferometric techniques, could provide empirical evidence for the quantum nature of gravity. We demonstrate that the geometrical configuration of an interferometer plays a decisive role in entanglement generation. Our analysis reveals that geometric optimization can substantially enhance the gravitationally induced phase shift, thereby improving the prospects for experimental detection of gravity-mediated entanglement. Moreover, we show that standard approximations employed in prior proposals may severely limit predictive accuracy and compromise experimental feasibility.
	\end{abstract}
	
	\maketitle
	
	\section{Introduction}
	\label{sec:intro}
	
	Quantum mechanics and general relativity are the two foundational theories of modern physics which together form the basis of our understanding of the physical world from microscopic particles to cosmic phenomena \cite{lawrie2012unified,buchbinder2021introduction,maleki2024cosmic}. Despite their successes, general relativity faces challenges at extreme energy scales, where singularities and breakdowns in its predictions highlight its limitations. The presence of singularities and the black hole information paradox---arising from quantum effects near black hole horizons and their subsequent evaporation \cite{hawking1974black,maleki2020speed}---indicate a need for a proper understanding of the quantum nature of gravity. Numerous approaches, including string theory and loop quantum gravity, aim to develop a quantum theory of gravity \cite{ashtekar2005gravity,kiefer2007quantum,rovelli2008loop}. However, experimental exploration of quantum gravity remains daunting due to the inaccessibility of the Planck scale \cite{alfaro2005quantum}.
	
	Alternative perspectives, such as viewing gravity as an emergent phenomenon rather than a fundamental interaction, have gained attraction \cite{padmanabhan2015emergent,verlinde2017emergent}. The need for empirical evidence to distinguish classical and quantum descriptions of gravity has spurred creative proposals. For instance, Richard Feynman's 1957 suggestion of a Stern-Gerlach-type experiment with spin superpositions linked gravitational interactions to quantum effects, highlighting the necessity of quantizing gravity \cite{cecile2011role}.
	
	Recently, advances in quantum information science and precision interferometry have revived the possibility of testing whether gravity must itself admit genuinely quantum features \cite{peres2004quantum,lunghi2013experimental,marletto2018quantum,haine2021searching,bian2025experimental}. In particular, a major line of research has focused on interferometric protocols in which gravity generates measurable phase shifts between spatially separated branches of massive quantum systems. Seminal proposals based on Stern--Gerlach and Mach--Zehnder-type interferometric settings showed that, if two neutral masses are each prepared in spatial superposition, their mutual Newtonian interaction can imprint path-dependent phases and, under suitable conditions, entangle the two systems \cite{bose2017spin,marletto2017gravitationally}. The observation of such gravitationally induced entanglement has therefore been widely discussed as a potential witness of the nonclassical, and possibly quantum, character of the gravitational interaction, provided the underlying assumptions of the argument are satisfied and no alternative classical mediation mechanism can account for the observed correlations \cite{bose2017spin,marletto2017gravitationally,hanif2024testing,torovs2021relative,howl2021non,avila2023quantum}.
	
	This perspective has stimulated a broad theoretical effort aimed at clarifying both the operational meaning and the experimental accessibility of gravity-mediated entanglement. Subsequent studies have refined the original proposals by analyzing relative degrees of freedom, noise sources, complementarity constraints, and alternative formulations of the witness protocol, thereby strengthening the conceptual framework within which these experiments are interpreted \cite{torovs2021relative,howl2021non,margalit2021realization,maleki2022complementarity,bose2025massive,marletto2025quantum}.
	
	In parallel, several works have investigated the observability of the entanglement signal, optimal parameter regimes, decoherence channels, and strategies for enhancing the gravitational phase, all of which are crucial for translating the original idea into a realistic laboratory test \cite{carney2021using,krisnanda2020observable,pedernales2022enhancing,maleki2022complementarity,ming2026unveiling,schut2023relaxation,weber2024bose,di2024bose,bose2509spin}. Notably, among alternative approaches in this direction, \cite{haine2021searching} developed a multi-parameter estimation scheme using ultracold atomic ensembles. This scheme shows how the quantum nature of the gravitational field leaves a detectable signature, and it takes advantage of the scalability and stability of many-particle systems. \cite{carney2021using} proposed an interactive quantum information sensing protocol, in which an atom--mechanical oscillator setup provides an indirect witness of entanglement generation that remains robust against typical thermal noise sources.
	
	These developments have reinforced the view that interferometric tests of gravity are no longer purely speculative thought experiments, but are progressively becoming experimentally structured programs.
	
	A central requirement in most of these proposals is the preparation and coherent manipulation of massive quantum systems in the approximate range of $10^{-14}$--$10^{-12}\,\mathrm{kg}$, where gravitationally generated phases may become appreciable while quantum coherence is still, in principle, maintainable \cite{bose2017spin,marletto2017gravitationally,schut2023relaxation,weber2024bose,di2024bose,bose2509spin}. Reaching this regime is highly nontrivial, since the desired signal competes with environmental decoherence, thermal noise, imperfect control over the quantum degree of freedom, and finite interferometric visibility. Nevertheless, recent experimental progress has reported important milestones, including Stern--Gerlach interferometry with massive particles, spatial superposition techniques, and coherent control of increasingly large systems \cite{margalit2021realization,keil2021stern,pedalino2026probing,arndt2014testing,gerlich2011quantum,bild2023schrodinger,inarrea2025schrodinger}. In this sense, the interferometric state can be viewed as a Schr\"odinger-cat state whose coherence renders tiny gravitational effects accessible to precision measurement \cite{maleki2018generating,yang2025hot}.
	
	Despite the promise of these interferometric schemes, substantial theoretical and practical challenges remain. In this work, we therefore revisit proposals for detecting gravitationally induced quantum entanglement and critically examine several commonly adopted approximations concerning the geometrical configuration of the interferometers. By developing exact solutions, we investigate how geometric factors influence experimental observables and identify optimized setups that significantly enhance gravitational phase accumulation. Our findings refine the theoretical framework underlying these experiments and provide practical strategies for improving their feasibility and reliability. Given that enhancing predictive accuracy is a crucial step toward realizing such experiments in the laboratory, this work addresses critical gaps in current methodology and establishes a more rigorous foundation for the experimental exploration of quantum gravity.
	
	\section{Experimental Setup}
	\label{sec:setup}
	
	The state space of a two-qubit system is represented as the product of the Hilbert spaces of the individual qubits, denoted by $H^{C}_1 \otimes H^{C}_2$, forming a 4-dimensional Hilbert space. Each qubit's Hilbert space is spanned by the orthogonal basis $\{\vert 0\rangle, \vert 1\rangle\}$, resulting in a two-qubit basis set $\{\vert 00\rangle, \vert 01\rangle, \vert 10\rangle, \vert 11\rangle\}$, where $\vert ij\rangle = \vert i\rangle \otimes \vert j\rangle$ for $i,j \in \{0,1\}$. Thus, the general pure state of a two-qubit system can be written as
	\begin{align}
		\vert \psi\rangle = \alpha_{0}\vert 00\rangle + \alpha_{1}\vert 01\rangle + \alpha_{2}\vert 10\rangle + \alpha_{3}\vert 11\rangle.
		\label{2state}
	\end{align}
	
\begin{figure}[ht]
		\centering
		\includegraphics[width=\columnwidth]{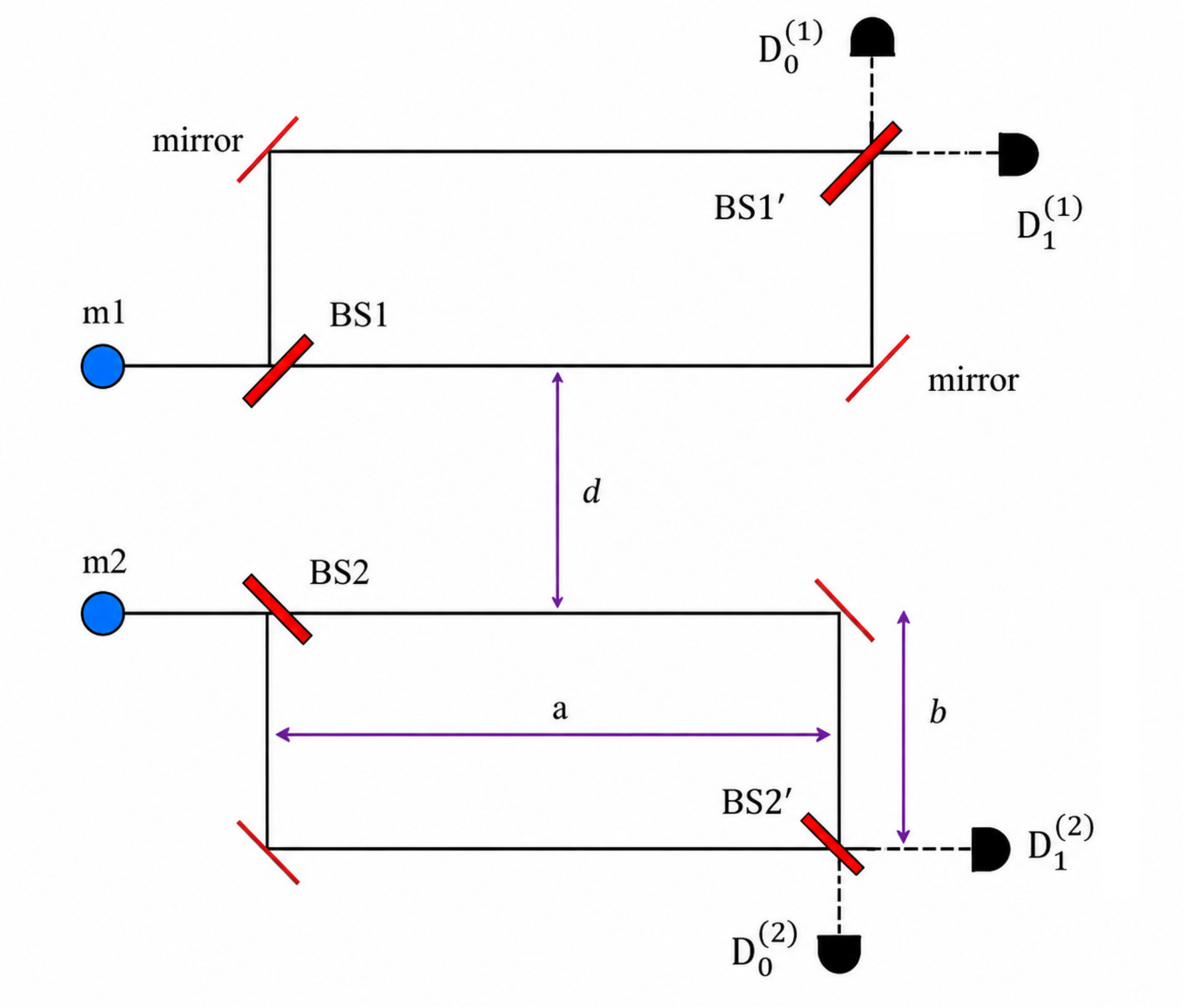}
		\caption{Experimental setup for generating gravitationally induced quantum entanglement using two Mach-Zehnder interferometers for the massive particles.}
		\label{fig1}
	\end{figure}

	Two-state systems, such as particle paths in Young's double-slit or Mach-Zehnder interferometers, can be represented within the two-qubit framework \cite{bose2017spin,marletto2017gravitationally}. To investigate the quantum aspects of gravity, we consider two massive particles, each passing through a Mach-Zehnder interferometer, as depicted in Fig.~\ref{fig1}, which has been investigated in recent literature \cite{marletto2017gravitationally,ming2026unveiling,weber2024bose,di2024bose,aimet2022gravity}. In this setup, the horizontal arm has a length of $a$, and the vertical arm has a length of $b$. The total path length that a particle travels before reaching the detector is given by $l = a + b$. We assume two identical Mach-Zehnder interferometers, separated by a distance $d$. The upper and lower paths in the interferometer are represented by the orthogonal states $\ket{u_i}$ and $\ket{d_i}$, corresponding to $\ket{1}$ and $\ket{0}$, respectively, as defined in Eq.~(\ref{2state}) for the $i$-th interferometer ($i = 1,2$).
	
	To explore the gravitationally induced entanglement, we consider two massive particles $m_1$ and $m_2$, each traversing an interferometer. The first beam splitter in the $i$-th interferometer ($i=1,2$) has reflectivity and transmissivity parameters $(r_i, t_i)$. The initial state after the first beam splitters is
	\begin{equation}
		\vert \psi(t=0)\rangle = ( r_1\vert u_1 \rangle +t_1\vert d_1\rangle) \otimes (t_2\vert u_2\rangle + r_2\vert d_2\rangle).
		\label{1ifnt}
	\end{equation}
	
	Gravitational interactions induce a phase shift in the quantum state of particles, as first demonstrated in the Colella-Overhauser-Werner (COW) experiment \cite{colella1975observation}
	\begin{align}
		\phi_{G} = -\int \frac{m \Phi}{\hbar} dt,
		\label{Gphase}
	\end{align}
	where $m$ is the particle mass, $\Phi$ is the gravitational potential, and $\hbar$ is the reduced Planck constant.
	
	In our setup, the gravitational field between two particles, each within its own interferometer and separated by a fixed distance, induces a phase shift in the quantum state \cite{bose2017spin,marletto2017gravitationally}. This configuration enables a detailed investigation of gravitationally induced quantum effects.
	
	In this setup, gravity induces a phase shift on the quantum states of the particles and effectively decouples from the system once the phase shift is established \cite{bose2017spin,marletto2017gravitationally}. As a result, the initial state of the system, $|\psi(t=0)\rangle$, evolves into the state
	\begin{align}
		\nonumber
		|\psi(t=T)\rangle &= t_1r_2 e^{i\phi_1}|d_1\rangle |d_2\rangle + t_1t_2 e^{i\phi_2}|d_1\rangle |u_2\rangle \\
		&\quad +r_1r_2 e^{i\phi_3}|u_1\rangle |d_2\rangle + r_1t_2 e^{i\phi_4}|u_1\rangle |u_2\rangle,
		\label{3ifnt}
	\end{align}
	just before entering the second beam splitter in each interferometer.
	
	In this context, it is argued that the lengths of the interferometers are significantly greater than their widths. Consequently, the interaction of the masses while traveling along the horizontal paths is taken into account as the main contribution to the phase accumulation. As a result, the accumulated phases are typically expressed as
	\begin{align}
		\phi_1 &=\phi_4= G\frac{m_1m_2T}{\hbar (d+b)},\\
		\phi_2&= G\frac{m_1m_2T}{\hbar d}, \\
		\phi_3&= G\frac{m_1m_2T}{\hbar (d+2b)}.
	\end{align}
	In the equations above, $G$ is the gravitational constant, and $T = l/v$ is the time duration the particles spend in the interferometer, where $v$ is the speed of the particle moving in the interferometer. Therefore, the phase accumulations along the vertical segments of the paths are often considered negligible in the literature. However, it is evident that for sufficiently small values of $b$ (small width), the paths do not separate far enough to accumulate a significantly large phase difference capable of generating entanglement. Thus, one is naturally led to consider non-negligible values of $b$, where the vertical phase contributions become appreciable. This raises the central question of our work as to what value of $b$ constitutes the optimal configuration for maximizing entanglement for a fixed total path length. In this work, we consider the case where the particle maintains a constant speed throughout its entire path. However, allowing the speed to vary based on the mass's location may lead to different conclusions.
	
	To this end, we consider the exact solution to the problem, incorporating contributions from the vertical paths. Taking these factors into account, the gravitationally induced phases $\phi_1$ and $\phi_4$ are determined to be
	\begin{align}
		\nonumber
		\phi_1 &= \phi_4 = G\frac{m_1m_2}{\hbar v} \times \\
		&\begin{cases}
			f(b,d) + f(b,d+b) + \frac{l-2b}{\sqrt{(d+b)^2+b^2}}, & \text{if $b \leq l/2$}, \\
			f(l-b,d) + f(l-b,d+3b-l) + g(b, d, l), & \text{otherwise},
		\end{cases}
		\label{Phi14}
	\end{align}
	where
	\begin{equation}
		f(x, y) = \frac{1}{\sqrt{2}} \ln \left( \frac{2x + y + \sqrt{2(x + y)^2 + 2x^2}}{y(1 + \sqrt{2})} \right),
		\label{fxy}
	\end{equation}
	and
	\begin{equation}
		g(b,d,l)= \frac{1}{2} \ln \left( \frac{d+3b-l+\sqrt{(l-b)^2+(d+3b-l)^2}}{d+l-b+\sqrt{(l-b)^2+(d+l-b)^2}} \right).
		\label{gbdl}
	\end{equation}
	Similarly, the gravitationally induced phases $\phi_2$ and $\phi_3$ are determined to be
	\begin{align}
		\nonumber
		\phi_2 &= G\frac{m_1m_2}{\hbar v} \left( \frac{1}{2} \ln \left( \frac{d+2b}{d} \right) + \frac{l-b}{d} \right), \\
		\phi_3 &= G\frac{m_1m_2}{\hbar v} \left( \frac{1}{2} \ln \left( \frac{d+2b}{d} \right) + \frac{l-b}{d+2b} \right).
	\end{align}
	The primary challenge in implementing this protocol lies in generating sufficiently large entanglement to make it experimentally detectable, a task anticipated to be highly demanding. To better understand this issue, we analyze the entanglement in the system through an appropriate entanglement measure.
	
	The amount of entanglement for a two-qubit quantum state $\rho$ can be quantified using the concurrence, defined based on the $R$ matrix as $R = \sqrt{\sqrt{\rho} \bar{\rho} \sqrt{\rho}}$, where $\bar{\rho} = (\sigma_y \otimes \sigma_y) \rho^\star (\sigma_y \otimes \sigma_y)$. The concurrence is determined from the eigenvalues of $R$, $\lambda_i$, arranged in decreasing order, using the formula \cite{Wootters1998}
	\begin{align}
		\mathcal{C} = \text{max}\{0, \lambda_0 - \lambda_1 - \lambda_2 - \lambda_3\}.
	\end{align}
	
	Therefore, from Eq.~(\ref{3ifnt}), the concurrence can be obtained as
	\begin{align}
		\mathcal{C} = 4(r_1 r_2 t_1 t_2) \vert \sin(\Delta\phi/2)\vert,
		\label{2C}
	\end{align}
	where $\Delta\phi = 2\phi_1 - \phi_2 - \phi_3$.
	
	The concurrence reaches its maximum when the beam splitter reflectivity parameters are set as $r_1 = r_2 = 1/\sqrt{2}$. In other words, to achieve maximum entanglement in the system, the optimal configuration is to use 50:50 beam splitters for all the beam splitters in the setup.
	
	Considering 50:50 beam splitters, the concurrence in the system reduces to
	\begin{align}
		\mathcal{C} =\vert \sin(\Delta\phi/2)\vert.
		\label{}
	\end{align}
	Evidently, this quantity varies between 0 and 1, as expected. The generated entanglement demonstrates an oscillatory behavior, where the extremum of the entanglement is reached at $\Delta\phi = k\pi$, with the maximum occurring for odd integers $k$, while no entanglement is generated for even integers $k$. Notably, for the observation of gravitationally induced entanglement, it may not be necessary to achieve maximum entanglement. A significant degree of entanglement may suffice to verify its presence in the system. For instance, $\mathcal{C} = 1/\sqrt{2} \approx 0.7$ can be achieved for $\Delta\phi = \pi/2$, which requires half of the accumulated phase that is needed to achieve $\mathcal{C} = 1$.
	
	Considering that in analyses related to the experimental feasibility of gravitationally induced entanglement, the factor $\kappa = \frac{G m_1 m_2}{\hbar v}$ is a decisive element, the primary objective has been to adjust the masses of the quantum particles ($m_1, m_2$) to be sufficiently large and the interaction time duration ($T = l/v$) to be long enough to accumulate significant phases in the system. Each of these objectives presents significant experimental challenges that must be addressed appropriately. However, the crucial role of the phase combination in Eq.~\ref{2C} and its dependence on the geometric configuration of the setup has been overlooked in prior studies.
	
	Now, we demonstrate that optimizing the geometric configuration of the setup can substantially enhance the induced phase, thereby improving the entanglement. Conversely, inappropriate geometric adjustments may diminish or even eliminate the induced entanglement. For this analysis, we assume that the total path traversed by the particles (the sum of the vertical and longitudinal arms) is fixed, as depicted in Fig.~\ref{fig1}, and for a fixed path length, we consider the optimal setting where the interferometer needs to be designed to maximize the entanglement.
	
	To this aim, one needs to optimize the vertical arm of the interferometers, $b$, relative to their total length, $l$. It is usually assumed that $l\gg b$, such that the contribution from the vertical segment can be ignored. Thus, the relevant factor to consider in the optimal setting is the ratio $b/l$.
	
	To analyze this, since $\kappa = \frac{G m_1 m_2}{\hbar v}$ is a common factor in all the phases, we consider the rescaled phase factor $\Delta\phi/\kappa$. Another relevant parameter in phase generation is the separation between the two interferometers. For the gravitational interaction between the masses to significantly enhance the induced phase, the interferometer separation $d$ must be relatively small compared to the interferometer arm length $l$. Thus, in experimental implementations, one would naturally aim for a small ratio $d/l$ to maximize the phase difference. However, the particles cannot be brought arbitrarily close. At sufficiently small separations, short-range interactions such as van der Waals forces dominate over the gravitational interaction. This sets a lower bound on the achievable separation $d$ \cite{van2020quantum}.
	
	\begin{figure}[t]
		\centering
		\includegraphics[width=\columnwidth]{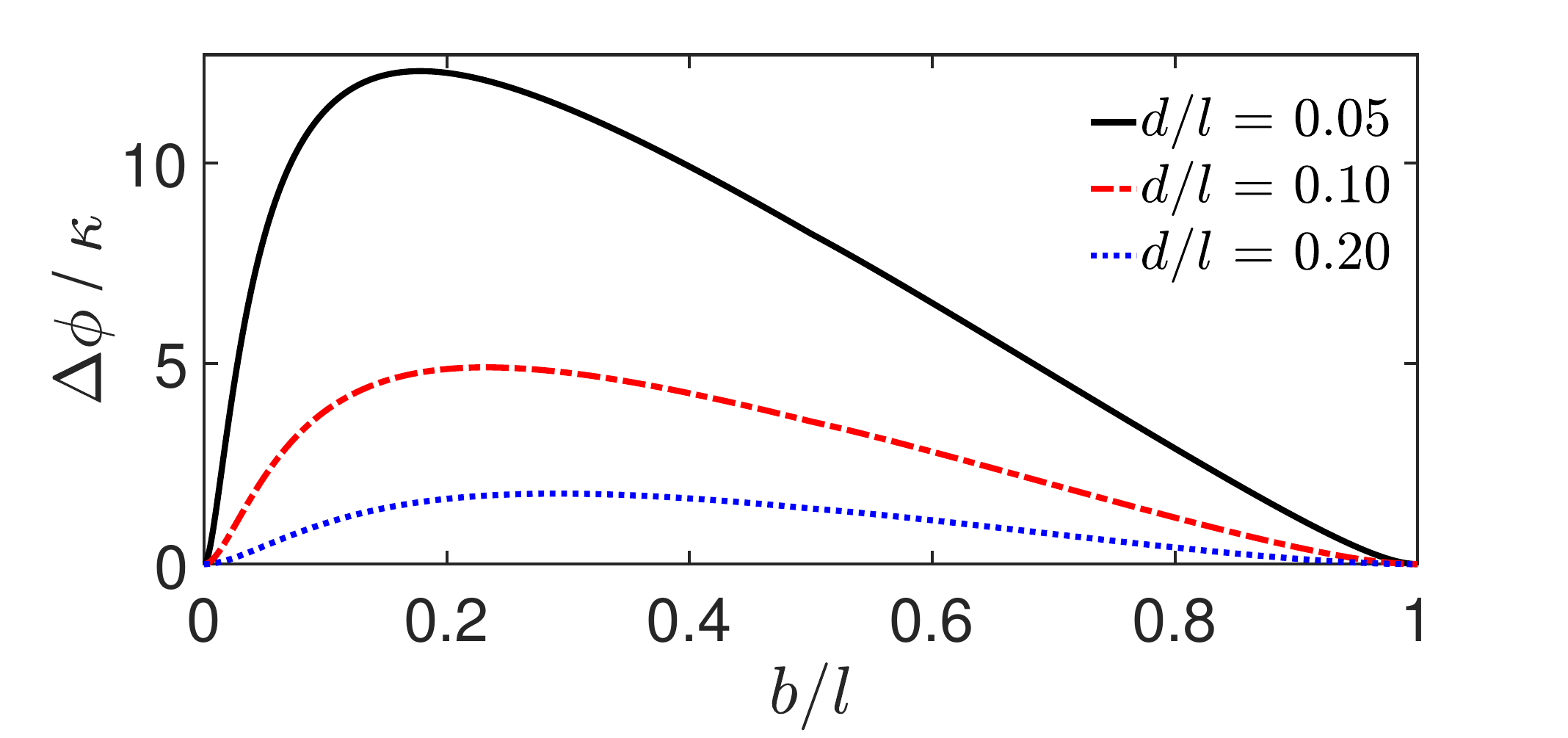}
		\caption{The induced phase as a function of the relative vertical arm length, $b/l$. The phase is rescaled to the factor $\kappa = G{m_1m_2}/(\hbar v)$. The separation between the two beam splitters is shown for three different values of $d/l$.}
		\label{fig1opt}
	\end{figure}
	
	\begin{figure}[t]
		\centering
		\includegraphics[width=\columnwidth]{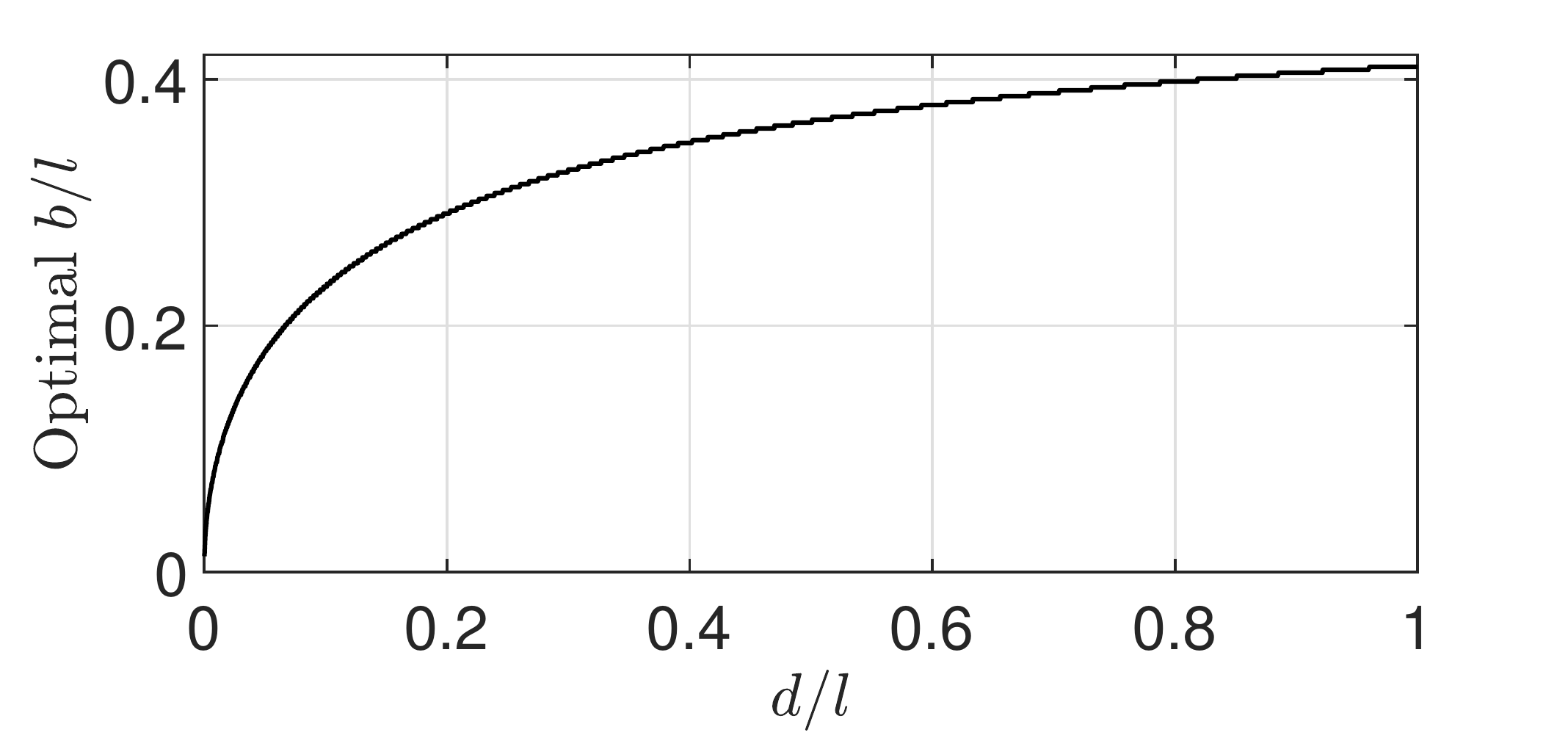}
		\caption{The optimal vertical arm length as a function of the distance between the two interferometers. Both lengths are rescaled to the total path of the interferometers, $l$.}
		\label{fig2opt}
	\end{figure}
	
	With this description, in Fig.~\ref{fig1opt}, we present the rescaled induced phase, $\Delta\phi/\kappa$, as a function of $b/l$. As seen from the figure, when the distance between the two interferometers is set to $d/l = 0.10$, the optimal configuration occurs at $b/l = 0.23$, resulting in a rescaled induced phase accumulation of about $4.9$. Note that the optimal setting does not happen at the negligible $b/l$ limit, which might seem a reasonable assumption at first encounter. Similarly, for $d/l = 0.05$, the optimal induced phase is achieved at $b/l = 0.18$, with a rescaled value of about $12.3$. This demonstrates that optimizing the vertical arm length within a specific range significantly enhances the induced phase. Also, we observe that the optimal vertical arm depends on the distance between the two interferometers. The optimal value can significantly affect the phase outcome. As can be seen from the figure, deviation from the optimal value can significantly affect the phase accumulation. Therefore, negligible values of $b/l\ll 1$ can result in significant challenges in the realization of this experiment. We also observe that, by increasing the distance between the interferometers, the phase accumulation becomes smaller. This is expected since the gravitational interaction between the interferometers becomes weaker.
	
	\begin{figure}[t]
		\centering
		\includegraphics[width=\columnwidth]{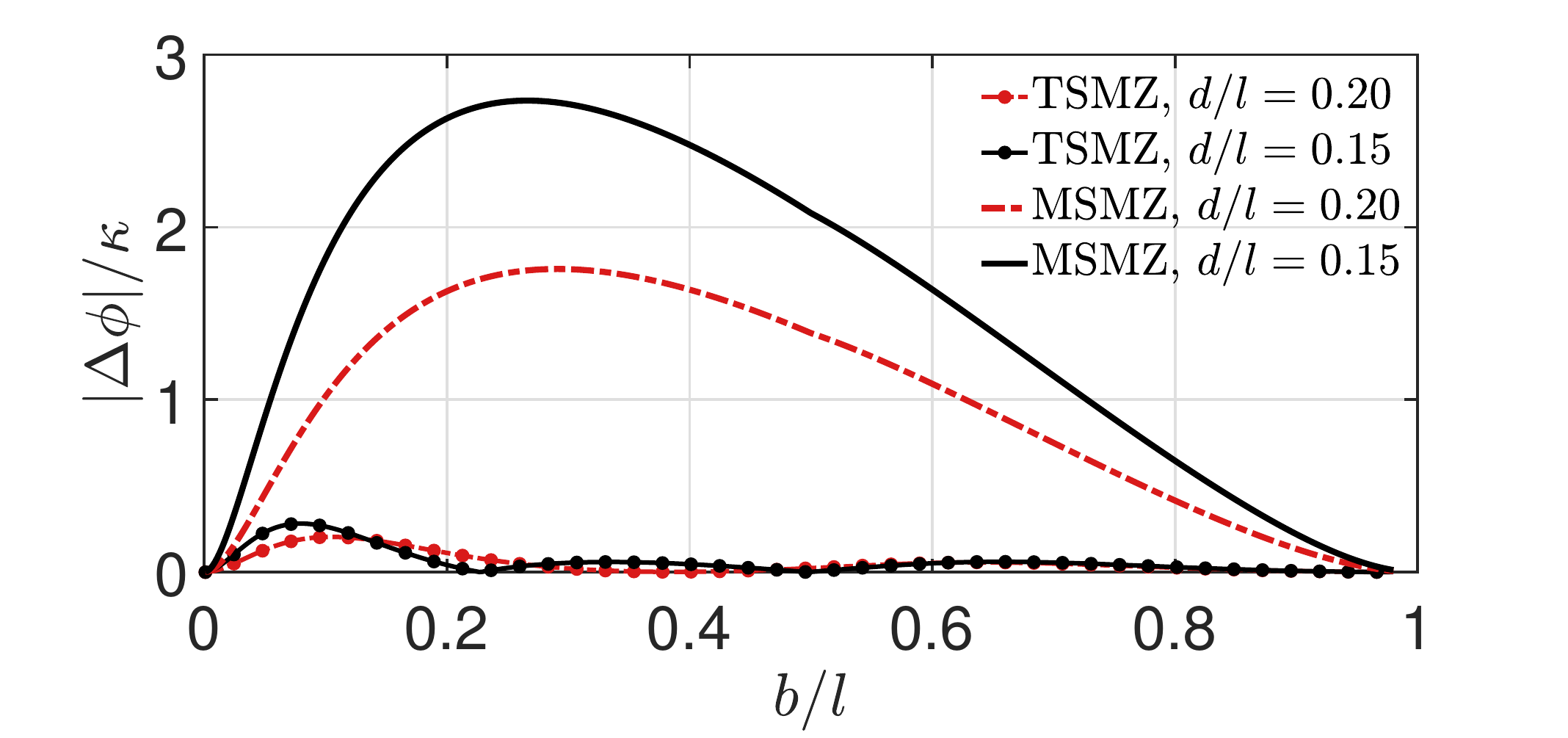}
		\caption{Rescaled phase $|\Delta\phi|/\kappa$ vs.\ $b/l$ for $d/l=0.20$
			(dash-dot) and $d/l=0.15$ (solid). Curves with markers:
			translation-symmetric Mach--Zehnder (TSMZ) setup of
			Ref.~\cite{marletto2017gravitationally}. Curves without markers:
			mirror-symmetric Mach--Zehnder (MSMZ) setup analyzed here. The MSMZ
			configuration yields a larger phase, with the effect being stronger at
			smaller $d/l$.}
		\label{figcompare}
	\end{figure}
	
	As another important result, Ref.~\cite{maleki2026role} showed that the originally proposed Mach--Zehnder setup of Ref.~\cite{marletto2017gravitationally} can yield null results at specific intermediate values of $b/l$, besides the edge limits. Here, we instead adopt a mirror-symmetric configuration. Figure~\ref{figcompare} compares the rescaled phase $|\Delta\phi|/\kappa$ of the translation-symmetric Mach--Zehnder (TSMZ) and mirror-symmetric Mach--Zehnder (MSMZ) setups investigated here for $d/l = 0.20$ and $0.15$. In both cases, the MSMZ setup results in a larger phase, especially for smaller $d/l$, which is more favorable for a real experiment. This further emphasizes the importance of the geometrical configuration in such setups.
	
	To analyze the relationship between the separation of the interferometers and their optimal geometrical configuration, we present in Fig.~\ref{fig2opt} the optimal values of $b/l$ as a function of the relative separation $d/l$. The results show that as the interferometers are brought closer together (smaller $d/l$), the optimal ratio $b/l$ also decreases. Therefore, in designing an experimental setup, one must carefully account for this dependence in order to achieve optimal sensitivity. As evident from Fig.~\ref{fig2opt}, for $d/l = 0.40$, the optimal value is $b/l = 0.35$, which is relatively large, indicating the importance of the vertical segment of the path.
	
	Therefore, to achieve a significant phase factor, the interferometers must be engineered such that their mutual separation is minimized (see Fig.~\ref{fig1opt}). Once the minimum achievable distance is established, the internal geometry of each interferometer should be tuned to the optimal ratio $b/l$ to maximize the phase accumulation.
	
\begin{figure}[ht]
		\centering
		\includegraphics[width=\columnwidth]{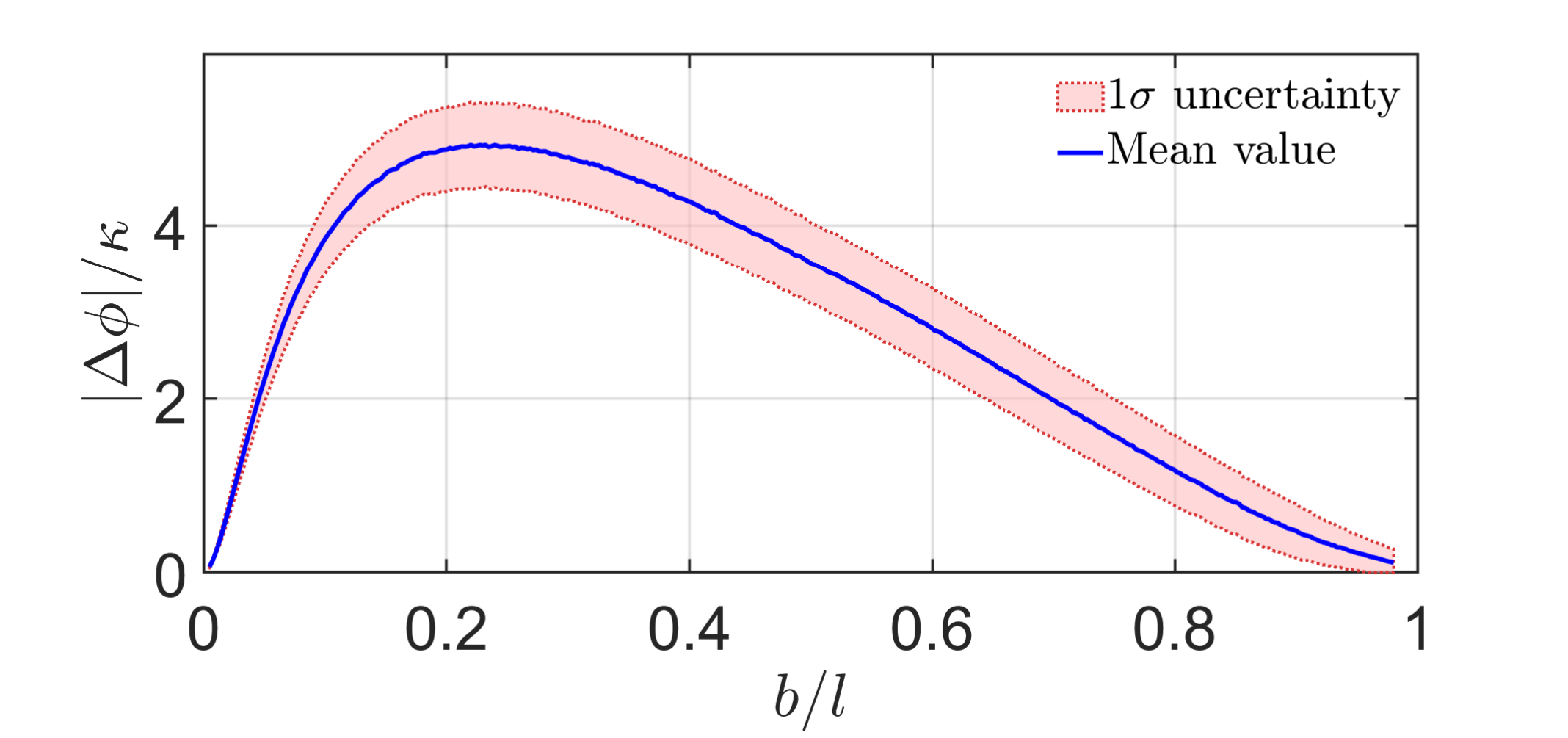}
		\caption{Absolute rescaled gravitational phase $|\Delta\phi|/\kappa$ vs. geometric ratio $b/l$. The solid curve shows the mean phase shift, and the red shaded area depicts the $1\sigma$ uncertainty band derived from $5\%$ uncertainties in $l$ and $d$. The absolute error is maximal at the signal peak, highlighting a fundamental trade-off between maximizing phase accumulation and maintaining tolerance to experimental uncertainties. Here we set $d/l=0.10$.}
		\label{PhaseERR}
	\end{figure}

	However, operating at this peak requires a careful assessment of experimental robustness, as the maximum signal strength often coincides with increased sensitivity to parameter fluctuations. This sensitivity is explored in Fig.~\ref{PhaseERR}, where the solid blue line represents the mean rescaled phase $|\Delta\phi|/\kappa$ obtained from a Monte Carlo simulation, and the shaded red region indicates the $1\sigma$ uncertainty band resulting from $5\%$ Gaussian fluctuations in the arm length $l$ and vertical separation $d$.
	
	A key observation is the non-uniform width of this error band. The uncertainty is suppressed as $b/\ell \to 0$ and $b/\ell \to 1$, regimes where the phase signal vanishes and the system becomes inherently robust against small geometric perturbations. In contrast, the uncertainty reaches its maximum near the optimal configuration ($b/\ell= 0.23$ for the case of $d/l=0.10$). Consequently, while the peak region offers the strongest signal for detecting gravity-induced entanglement, it also imposes the most stringent requirements on experimental precision and fabrication tolerances. To mitigate this uncertainty, one can employ a statistical approach by repeating the measurement over $N$ independent trials. The precision of the phase estimation then improves according to the standard scaling law of $1/\sqrt{N}$, allowing for higher precision in the optimal regime.
	
	\section{Measurement of the Entanglement in Experiment}
	\label{Bell}
	
	Central to our discussion above was the quantification of entanglement through concurrence. A reasonable question might arise regarding the practical measurement of entanglement via concurrence. Since concurrence is not related to an observable quantity, there is no direct way to measure it. In other words, no Hermitian operator can be attributed to concurrence. This raises the question of how to practically quantify entanglement through concurrence.
	
	In fact, since the state of the two interferometers results in a two-qubit pure state, it turns out that concurrence can be analyzed using the Bell inequality \cite{bell1987einstein}.
	
	To clarify this, we discuss the relation between the Bell inequality and concurrence, which provides an experimentally testable approach to quantum entanglement. To this aim, we utilize a specific version of the Bell inequality, the so-called CHSH inequality which is given by \cite{clauser1969proposed,horodecki1995violating}
	\begin{equation}
		\hat{B}=\vec{a} \cdot \vec{\sigma} \otimes\left(\vec{b}+\vec{b^{\prime}}\right) \cdot \vec{\sigma}+\vec{a^{\prime}} \cdot \vec{\sigma} \otimes\left(\vec{b}-\vec{b^{\prime}}\right) \cdot \vec{\sigma},
	\end{equation}
	where $\vec{a}, \vec{a^{\prime}}, \vec{b}, \vec{b^{\prime}}$ are unit vectors. Using the relations defined above, the CHSH inequality for a given quantum state described by the density matrix $\rho$ is expressed as \cite{clauser1969proposed,horodecki1995violating}
	\begin{equation}
		|\langle\hat{B}\rangle|=|\operatorname{tr}(\rho \hat{B})| \leq 2.
	\end{equation}
	
	A violation of this inequality, i.e., $|\langle \hat{B} \rangle| > 2$, signifies the presence of quantum correlations within a quantum state. Thus, it offers a compelling experimental approach to probing the quantum nature of gravity.
	
	An intriguing connection exists between entanglement and the maximum value of the Bell parameter, expressed as \cite{ghosh2001mixedness,verstraete2002entanglement}
	\begin{align}\label{bell}
		{B}_{\text{max}}=2\sqrt{1+\mathcal{C}^{2}}.
	\end{align}
	Hence, from Eq.~\ref{2C}, the maximum value of the Bell parameter is given by
	\begin{align}
		\label{bell2}
		B_{\text{max}} = 2 \sqrt{ 1 + 16 ({r_1}{r_2}{t_1}{t_2})^2 \sin^2 \left( \frac{\Delta \phi}{2} \right) }.
	\end{align}
	
	Therefore, concurrence can be achieved from the measurement of the Bell inequality, in the setup of this work.
	
	\begin{figure}[ht]
		\centering
		\includegraphics[width=\columnwidth]{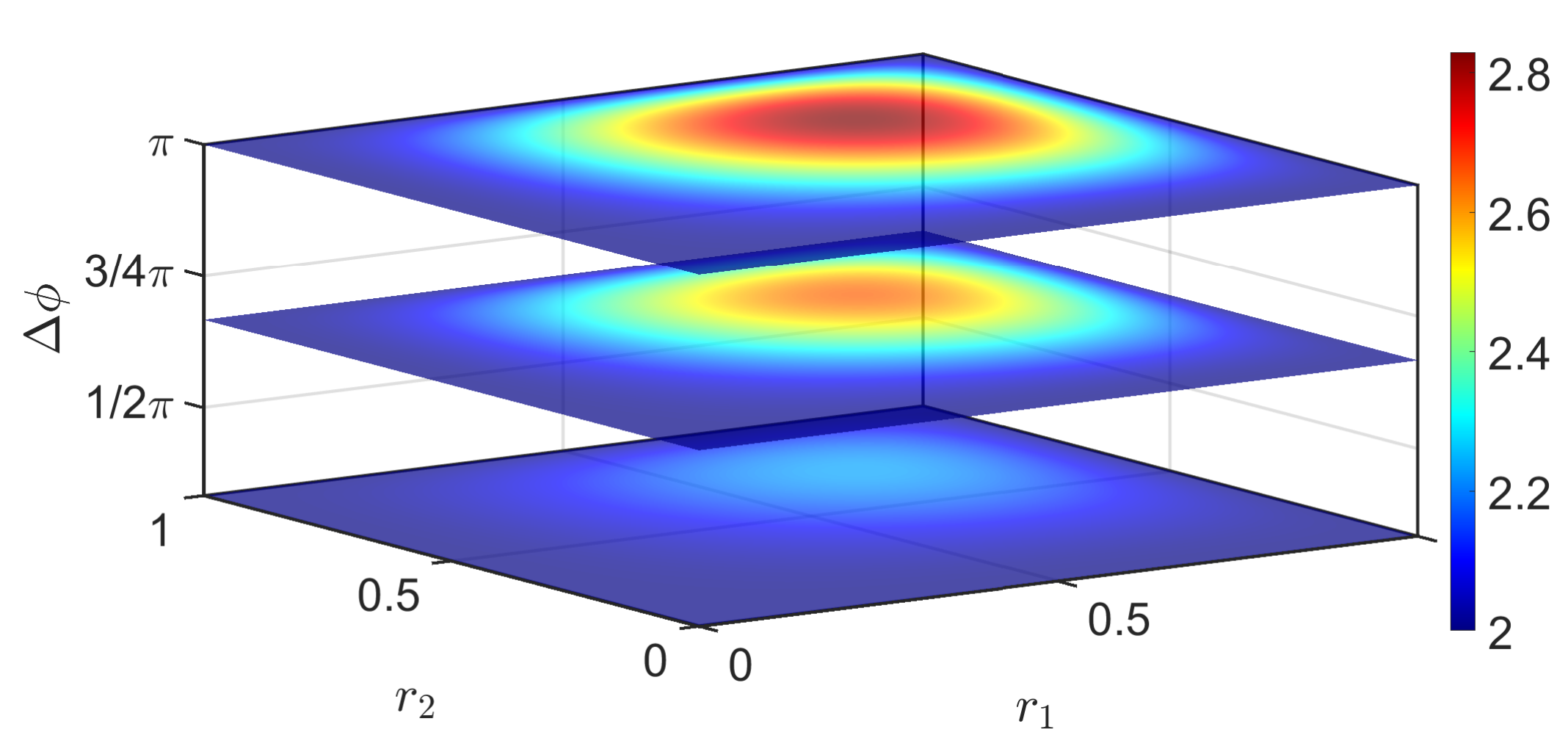}
		\caption{Maximum of the Bell inequality as a function of the reflectivity of two beam splitters. The plot demonstrates that optimal violation occurs when both beam splitters are set to 50:50. The z-axis represents the phase difference $\Delta\phi$, while the x- and y-axes correspond to the reflectivity parameters $r_1$ and $r_2$, respectively. The color scale indicates the extent of the violation.}
		\label{BmaxDplot}
	\end{figure}
	
	The plot shown in Fig.~\ref{BmaxDplot} illustrates the maximum of the Bell inequality as a function of the reflectivity parameters of two beam splitters and the induced phase difference. It is evident from the figure that the violation reaches its peak when the phase difference between the input beams, $\Delta\phi$, is set to $\pi$. However, more importantly, the optimal violation is achieved when the reflectivity of both beam splitters is set to 50:50. This configuration leads to the most efficient results, ensuring the highest violation of the Bell inequality.
	
	Therefore, in the context of testing quantum gravity, in interferometry experiments involving entanglement between two masses, the choice of beam splitter reflectivity and geometrical configuration can maximize the violation of the Bell inequality, which is essential for detecting quantum mechanical effects in interferometry experiments designed to probe quantum gravity using entanglement-induced interference.
	
	\section{Output State and Detection Probabilities}
	
	Including the second beam splitter in the interferometers recombines the two possible paths of each particle and therefore allows phase-dependent interference between the corresponding probability amplitudes. The amplitudes associated with transmission and reflection at the beam splitters, together with the phase differences accumulated along the interferometer arms, interfere when the paths meet again at the second beam splitters. As a consequence, the joint detection probabilities at the detectors $D_0^{(i)}$ and $D_1^{(i)}$ depend on both the beam-splitter coefficients (see Fig.~\ref{fig1}).
	
	For simplicity we define the following coefficients
	\begin{equation}
		A_{dd}=t_1r_2e^{i\phi_1}, \qquad
		A_{du}=t_1t_2e^{i\phi_2},
	\end{equation}
	and
	\begin{equation}
		A_{ud}=r_1r_2e^{i\phi_3}, \qquad
		A_{uu}=r_1t_2e^{i\phi_4}.
	\end{equation}
	
	Therefore, the two-particle state in Eq.~\ref{3ifnt}, just before the second beam splitters can be written as
	\begin{equation}
		\begin{aligned}
			|\psi(T)\rangle
			&=
			A_{dd}|d_1d_2\rangle
			+
			A_{du}|d_1u_2\rangle
			+
			A_{ud}|u_1d_2\rangle
			+
			A_{uu}|u_1u_2\rangle ,
		\end{aligned}
	\end{equation}
	
	For a general second beam splitter in the $i$-th interferometer, we use the transformation
	\begin{equation}
		|u_i\rangle
		\rightarrow
		r_i'|D_0^{(i)}\rangle
		+
		t_i'|D_1^{(i)}\rangle ,
	\end{equation}
	and
	\begin{equation}
		|d_i\rangle
		\rightarrow
		t_i'|D_0^{(i)}\rangle
		-
		r_i'|D_1^{(i)}\rangle ,
	\end{equation}
	where $r_i'$ and $t_i'$ are the reflection and transmission amplitudes of the second beam splitter. Applying this transformation to each of the four path states, the output state becomes
	\begin{equation}
		\begin{aligned}
			|\psi_{\mathrm{out}}\rangle
			&=
			B_{00}|D_0^{(1)}D_0^{(2)}\rangle
			+
			B_{01}|D_0^{(1)}D_1^{(2)}\rangle  \\
			&\quad+
			B_{10}|D_1^{(1)}D_0^{(2)}\rangle
			+
			B_{11}|D_1^{(1)}D_1^{(2)}\rangle ,
		\end{aligned}
	\end{equation}
	where
	\begin{equation}
		\begin{aligned}
			B_{00}
			&=
			A_{dd}t_1't_2'
			+
			A_{du}t_1'r_2'
			+
			A_{ud}r_1't_2'
			+
			A_{uu}r_1'r_2',
		\end{aligned}
	\end{equation}
	\begin{equation}
		\begin{aligned}
			B_{01}
			&=
			-A_{dd}t_1'r_2'
			+
			A_{du}t_1't_2'
			-
			A_{ud}r_1'r_2'
			+
			A_{uu}r_1't_2',
		\end{aligned}
	\end{equation}
	\begin{equation}
		\begin{aligned}
			B_{10}
			&=
			-A_{dd}r_1't_2'
			-
			A_{du}r_1'r_2'
			+
			A_{ud}t_1't_2'
			+
			A_{uu}t_1'r_2',
		\end{aligned}
	\end{equation}
	\begin{equation}
		\begin{aligned}
			B_{11}
			&=
			A_{dd}r_1'r_2'
			-
			A_{du}r_1't_2'
			-
			A_{ud}t_1'r_2'
			+
			A_{uu}t_1't_2'.
		\end{aligned}
	\end{equation}
	
	Thus, the output probabilities are
	\begin{equation}
		P_{00}=|B_{00}|^2, \qquad
		P_{01}=|B_{01}|^2,
	\end{equation}
	and
	\begin{equation}
		P_{10}=|B_{10}|^2, \qquad
		P_{11}=|B_{11}|^2.
	\end{equation}
	The joint detection probabilities for the various outcomes following the second set of beam splitters are defined as follows: $P_{00}$ represents the case where both particles are detected at $D_0$, while $P_{11}$ denotes the event where both are detected at $D_1$. The cross-terms $P_{01}$ and $P_{10}$ correspond to the outcomes where particle 1 is detected at $D_0$ (with particle 2 at $D_1$) and particle 1 is detected at $D_1$ (with particle 2 at $D_0$), respectively.
	
	Explicitly, the probability for the first particle detected by $D_0^{(1)}$ outcome is
	\begin{equation}
		\begin{aligned}
			P_{0}^{1}
			&=P_{00}+P_{01}= t_1'^{2} t_1^{2} + r_1'^{2} r_1^{2} \\
			&\quad + 2\, t_1' r_1' t_1 r_1
			\big( r_2^{2}\cos(\phi_1 - \phi_3) \\
			&\quad + t_2^{2}\cos(\phi_2 - \phi_1) \big)
		\end{aligned}
	\end{equation}
	while the probability for the $D_1^{(1)}$ outcome is
	\begin{equation}
		\begin{aligned}
			P_{1}^{1}
			&=P_{10}+P_{11}
			= r_1'^{2} t_1^{2} + t_1'^{2} r_1^{2} \\
			&\quad - 2\, t_1' r_1' t_1 r_1
			\big( r_2^{2}\cos(\phi_1 - \phi_3) \\
			&\quad + t_2^{2}\cos(\phi_2 - \phi_1) \big)
		\end{aligned}
	\end{equation}
	For balanced (50:50) beam splitters, the joint detection probabilities are given by
	\begin{align}
		\nonumber
		P_{00} &= \frac{1}{8} \big[ 3 + 2\cos(\phi_2 - \phi_1) \\
		&\quad + 2\cos(\phi_3 - \phi_1) + \cos(\phi_2 - \phi_3) \big], \\
		P_{01} &= P_{10} = \frac{1}{4} \sin^2 \left( \frac{\phi_2 - \phi_3}{2} \right), \\
		\nonumber
		P_{11} &= \frac{1}{8} \big[ 3 - 2\cos(\phi_2 - \phi_1) \\
		&\quad - 2\cos(\phi_3 - \phi_1) + \cos(\phi_2 - \phi_3) \big].
	\end{align}
	In the non-interacting limit (i.e., $\phi_{i} = 0$), the particles undergo constructive interference, resulting in a joint detection at $D_0$ with unit probability ($P_{00} = 1$). However, the introduction of gravitational interaction induces relative phases that can lead to entanglement between the two masses. In the regime of maximal entanglement, the probability $P_{00}$ is reduced and modulated within the range $1/4 \leq P_{00} \leq 1/2$.
	
	\section{Summary and Conclusion}
	\label{Conclusion}
	
	In this work, we revisited the theoretical foundation and experimental feasibility of detecting gravitationally induced quantum entanglement in a two-mass interferometric setup. Using a general two-qubit representation, we systematically analyzed the evolution of the quantum state under gravitational interaction and derived exact analytical expressions for the induced phases. These phases, arising from the mutual gravitational coupling between the two masses, play a central role in the generation of entanglement.
	
	By examining a realistic interferometric configuration without relying on commonly used approximations, we obtained precise expressions for the gravitationally induced phases. Our analysis showed that neglecting the vertical arms of the interferometer can lead to significant inaccuracies. This result highlights the importance of treating the full geometry of the setup in order to obtain reliable predictions for the performance of such experiments.
	
	A central outcome of our study is that optimizing the geometric configuration of the interferometer, especially the vertical arm length, can substantially increase the gravitationally induced phase and thereby enhance the achievable entanglement. We also demonstrated that the mirror-symmetric Mach--Zehnder configuration yields a larger induced phase, with the advantage being most pronounced at small separations $d/l$, compared with the translation-symmetric Mach--Zehnder setup. This demonstrates that geometric optimization offers a practical and cost-effective strategy for improving the sensitivity of proposed tests of gravity-mediated entanglement.
	
	It is also important to clarify why the experimental observation of gravitationally induced entanglement is particularly difficult. The main challenge is that one must generate a detectable gravitational phase shift, typically of order unity, while operating in a regime where gravity remains the dominant source of correlation between the masses. Achieving this goal requires simultaneously overcoming several major obstacles. First, the masses must be sufficiently large to produce an appreciable gravitational interaction, yet they must also be prepared and maintained in coherent spatial superposition states, which becomes increasingly demanding as the mass grows. Second, the two systems must be brought close enough to enhance the gravitational coupling, often at micrometer-scale separations, while carefully suppressing much stronger non-gravitational backgrounds such as electrostatic or Casimir-type interactions. This in turn requires sophisticated shielding and isolation techniques, including conductive screening layers and precise control over the experimental environment. Third, quantum coherence must be preserved over the entire interaction time, which generally necessitates extreme control of decoherence sources, such as collisions, thermal radiation, vibrational noise, and internal dissipation, often under cryogenic conditions.
	
	These combined requirements place gravitationally induced entanglement experiments at the frontier of current quantum control capabilities. In this context, investigating the optimum geometry of the interferometer becomes especially valuable, since it provides a comparatively low-cost and experimentally accessible route to enhancing the gravitational phase in interferometric setups. Our main goal in this work has been to demonstrate the importance of this geometric factor in the interferometric realization of such systems. We therefore expect that the present analysis can serve as useful guidance for the design of future experiments aimed at probing the quantum nature of gravity.
	
	\bibliography{library}
	
\end{document}